\documentclass[aps,prd,twocolumn,superscriptaddress,letterpaper,floatfix,nofootinbib,showpacs,preprintnumbers]{revtex4}
\usepackage{epsfig,units,color,ulem}
\newcommand{\<}{\ensuremath{<\!}}

\renewcommand{\>}{\ensuremath{\!>}}
\begin{document}
%\pagewiselinenumbers
%\preprint{FERMILAB-PUB-XX-YYY-Z}

\title{Observation of muon intensity variations by season with the 
MINOS Near Detector}



% \author{J.~K. de~Jong}
%\email[]{jeffrey.dejong@physics.ox.ac.uk}
%\collaboration{The MINOS Collaboration}

\newcommand{\Berkeley}{Lawrence Berkeley National Laboratory, Berkeley, California, 94720 USA}
\newcommand{\Cambridge}{Cavendish Laboratory, University of Cambridge, Madingley Road, Cambridge CB3 0HE, United Kingdom}
\newcommand{\Cincinnati}{Department of Physics, University of Cincinnati, Cincinnati, Ohio 45221, USA}
\newcommand{\FNAL}{Fermi National Accelerator Laboratory, Batavia, Illinois 60510, USA}
\newcommand{\RAL}{Rutherford Appleton Laboratory, Science and TechnologiesFacilities Council, Didcot, OX11 0QX, United Kingdom}
\newcommand{\UCL}{Department of Physics and Astronomy, University College London, Gower Street, London WC1E 6BT, United Kingdom}
\newcommand{\Caltech}{Lauritsen Laboratory, California Institute of Technology, Pasadena, California 91125, USA}
\newcommand{\Alabama}{Department of Physics and Astronomy, University of Alabama, Tuscaloosa, Alabama 35487, USA}
\newcommand{\ANL}{Argonne National Laboratory, Argonne, Illinois 60439, USA}
\newcommand{\Athens}{Department of Physics, University of Athens, GR-15771 Athens, Greece}
\newcommand{\NTUAthens}{Department of Physics, National Tech. University of Athens, GR-15780 Athens, Greece}
\newcommand{\Benedictine}{Physics Department, Benedictine University, Lisle, Illinois 60532, USA}
\newcommand{\BNL}{Brookhaven National Laboratory, Upton, New York 11973, USA}
\newcommand{\CdF}{APC -- Universit\'{e} Paris 7 Denis Diderot, 10, rue Alice Domon et L\'{e}onie Duquet, F-75205 Paris Cedex 13, France}
\newcommand{\Cleveland}{Cleveland Clinic, Cleveland, Ohio 44195, USA}
\newcommand{\Delhi}{Department of Physics \& Astrophysics, University of Delhi, Delhi 110007, India}
\newcommand{\GEHealth}{GE Healthcare, Florence South Carolina 29501, USA}
\newcommand{\Harvard}{Department of Physics, Harvard University, Cambridge, Massachusetts 02138, USA}
\newcommand{\HolyCross}{Holy Cross College, Notre Dame, Indiana 46556, USA}
\newcommand{\Houston}{Department of Physics, University of Houston, Houston, Texas 77204, USA}
\newcommand{\IIT}{Department of Physics, Illinois Institute of Technology, Chicago, Illinois 60616, USA}
\newcommand{\Iowa}{Department of Physics and Astronomy, Iowa State University, Ames, Iowa 50011 USA}
\newcommand{\Indiana}{Indiana University, Bloomington, Indiana 47405, USA}
\newcommand{\ITEP}{High Energy Experimental Physics Department, ITEP, B. Cheremushkinskaya, 25, 117218 Moscow, Russia}
\newcommand{\JMU}{Physics Department, James Madison University, Harrisonburg, Virginia 22807, USA}
\newcommand{\LASL}{Nuclear Nonproliferation Division, Threat Reduction Directorate, Los Alamos National Laboratory, Los Alamos, New Mexico 87545, USA}
\newcommand{\Lebedev}{Nuclear Physics Department, Lebedev Physical Institute, Leninsky Prospect 53, 119991 Moscow, Russia}
\newcommand{\LLL}{Lawrence Livermore National Laboratory, Livermore, California 94550, USA}
\newcommand{\LosAlamos}{Los Alamos National Laboratory, Los Alamos, New Mexico 87545, USA}
\newcommand{\Manchester}{School of Physics and Astronomy, University of Manchester, Oxford Road, Manchester M13 9PL, United Kingdom}
\newcommand{\MIT}{Lincoln Laboratory, Massachusetts Institute of Technology, Lexington, Massachusetts 02420, USA}
\newcommand{\Minnesota}{University of Minnesota, Minneapolis, Minnesota 55455, USA}
\newcommand{\Crookston}{Math, Science and Technology Department, University of Minnesota -- Crookston, Crookston, Minnesota 56716, USA}
\newcommand{\Duluth}{Department of Physics, University of Minnesota Duluth, Duluth, Minnesota 55812, USA}
\newcommand{\Ohio}{Center for Cosmology and Astro Particle Physics, Ohio State University, Columbus, Ohio 43210 USA}
\newcommand{\Otterbein}{Otterbein College, Westerville, Ohio 43081, USA}
\newcommand{\Oxford}{Subdepartment of Particle Physics, University of Oxford, Oxford OX1 3RH, United Kingdom}
\newcommand{\PennState}{Department of Physics, Pennsylvania State University, State College, Pennsylvania 16802, USA}
\newcommand{\PennU}{Department of Physics and Astronomy, University of Pennsylvania, Philadelphia, Pennsylvania 19104, USA}
\newcommand{\Pittsburgh}{Department of Physics and Astronomy, University of Pittsburgh, Pittsburgh, Pennsylvania 15260, USA}
\newcommand{\IHEP}{Institute for High Energy Physics, Protvino, Moscow Region RU-140284, Russia}
\newcommand{\Rochester}{Department of Physics and Astronomy, University of Rochester, New York 14627 USA}
\newcommand{\RoyalH}{Physics Department, Royal Holloway, University of London, Egham, Surrey, TW20 0EX, United Kingdom}
\newcommand{\Carolina}{Department of Physics and Astronomy, University of South Carolina, Columbia, South Carolina 29208, USA}
\newcommand{\SLAC}{Stanford Linear Accelerator Center, Stanford, California 94309, USA}
\newcommand{\Stanford}{Department of Physics, Stanford University, Stanford, California 94305, USA}
\newcommand{\StJohnFisher}{Physics Department, St. John Fisher College, Rochester, New York 14618 USA}
\newcommand{\Sussex}{Department of Physics and Astronomy, University of Sussex, Falmer, Brighton BN1 9QH, United Kingdom}
\newcommand{\TexasAM}{Physics Department, Texas A\&M University, College Station, Texas 77843, USA}
\newcommand{\Texas}{Department of Physics, University of Texas at Austin, 1 University Station C1600, Austin, Texas 78712, USA}
\newcommand{\TechX}{Tech-X Corporation, Boulder, Colorado 80303, USA}
\newcommand{\Tufts}{Physics Department, Tufts University, Medford, Massachusetts 02155, USA}
\newcommand{\UNICAMP}{Universidade Estadual de Campinas, IFGW-UNICAMP, CP 6165, 13083-970, Campinas, SP, Brazil}
\newcommand{\UFG}{Instituto de F\'{i}sica, Universidade Federal de Goi\'{a}s, CP 131, 74001-970, Goi\^{a}nia, GO, Brazil}
\newcommand{\USP}{Instituto de F\'{i}sica, Universidade de S\~{a}o Paulo,  CP 66318, 05315-970, S\~{a}o Paulo, SP, Brazil}
\newcommand{\Warsaw}{Department of Physics, University of Warsaw, Ho\.{z}a 69, PL-00-681 Warsaw, Poland}
\newcommand{\Washington}{Physics Department, Western Washington University, Bellingham, Washington 98225, USA}
\newcommand{\WandM}{Department of Physics, College of William \& Mary, Williamsburg, Virginia 23187, USA}
\newcommand{\Wisconsin}{Physics Department, University of Wisconsin, Madison, Wisconsin 53706, USA}
\newcommand{\deceased}{Deceased.}

\affiliation{\ANL}
\affiliation{\Athens}
%\affiliation{\Benedictine}
\affiliation{\BNL}
\affiliation{\Caltech}
\affiliation{\Cambridge}
\affiliation{\UNICAMP}
%\affiliation{\CdF}
\affiliation{\Cincinnati}
\affiliation{\FNAL}
\affiliation{\UFG}
\affiliation{\Harvard}
\affiliation{\HolyCross}
\affiliation{\Houston}
\affiliation{\IIT}
\affiliation{\Indiana}
\affiliation{\Iowa}
%\affiliation{\IHEP}
%\affiliation{\ITEP}
%\affiliation{\JMU}
%\affiliation{\Lebedev}
%\affiliation{\LLL}
\affiliation{\UCL}
\affiliation{\Manchester}
\affiliation{\Minnesota}
\affiliation{\Duluth}
\affiliation{\Otterbein}
\affiliation{\Oxford}
\affiliation{\Pittsburgh}
\affiliation{\RAL}
\affiliation{\USP}
\affiliation{\Carolina}
\affiliation{\Stanford}
\affiliation{\Sussex}
\affiliation{\TexasAM}
\affiliation{\Texas}
\affiliation{\Tufts}
\affiliation{\Warsaw}
%\affiliation{\Washington}
\affiliation{\WandM}
%\affiliation{\Wisconsin}

\author{P.~Adamson}
\affiliation{\FNAL}
%\affiliation{\UCL}
%\affiliation{\Sussex}

%\author{C.~Andreopoulos}
%\affiliation{\RAL}
%\affiliation{\Athens}

\author{I.~Anghel}
\affiliation{\Iowa}
\affiliation{\ANL}

%\author{K.~E.~Arms}
%\affiliation{\Minnesota}

%\author{R.~Armstrong}
%\affiliation{\Indiana}

\author{A.~Aurisano}
\affiliation{\Cincinnati}

%\author{T.~H.~Fields}
%\affiliation{\ANL}

%\author{D.~J.~Auty}
%\affiliation{\Sussex}

%\author{S.~Avvakumov}
%\affiliation{\Stanford}

%\author{D.~S.~Ayres}
%\affiliation{\ANL}

%\author{C.~Backhouse}
%\affiliation{\Oxford}

%\author{B.~Baller}
%\affiliation{\FNAL}

%\author{B.~Barish}
%\affiliation{\Caltech}

%\author{P.~D.~Barnes~Jr.}
%\affiliation{\LLL}

\author{G.~Barr}
\affiliation{\Oxford}

%\author{W.~L.~Barrett}
%\affiliation{\Washington}

%\author{E.~Beall}
%\altaffiliation[Now at\ ]{\Cleveland .}
%\affiliation{\ANL}
%\affiliation{\Minnesota}

%\author{B.~R.~Becker}
%\affiliation{\Minnesota}

%\author{A.~Belias}
%\affiliation{\RAL}

%\author{R.~H.~Bernstein}
%\affiliation{\FNAL}

%\author{M.~Betancourt}
%\affiliation{\Minnesota}

%\author{D.~Bhattacharya}
%\affiliation{\Pittsburgh}

%\author{M.~Bhattarai}
%\affiliation{\Texas}
%\affiliation{\Duluth}

\author{M.~Bishai}
\affiliation{\BNL}

\author{A.~Blake}
\affiliation{\Cambridge}

%\author{B.~Bock}
%\affiliation{\Duluth}

\author{G.~J.~Bock}
\affiliation{\FNAL}

%\author{D.~J.~Boehnlein}
%\affiliation{\FNAL}

\author{D.~Bogert}
\affiliation{\FNAL}

%\author{P.~M.~Border}
%\affiliation{\Minnesota}

%\author{C.~Bower}
%\affiliation{\Indiana}

%\author{E.~Buckley-Geer}
%\affiliation{\FNAL}

\author{S.~V.~Cao}
\affiliation{\Texas}

\author{C.~M.~Castromonte}
\affiliation{\UFG}

%\author{S.~Cavanaugh}
%\affiliation{\Harvard}

%\author{J.~D.~Chapman}
%\affiliation{\Cambridge}

%\author{D.~Cherdack}
%\affiliation{\Tufts}

% ...

\collaboration{The MINOS Collaboration}
\noaffiliation

%\thanks{}
% \affiliation{University of Oxford}

%\collaboration{The MINOS Collaboration}

\date{\today}

\begin{abstract}
A sample of 1.53$\times$10$^{9}$
cosmic-ray-induced single muon 
events has been recorded at 225 meters-water-equivalent 
using the MINOS Near Detector.  The underground muon rate 
is observed to be highly correlated with the effective 
atmospheric temperature. The coefficient $\alpha_{T}$, relating 
the change in the muon rate to the change in the vertical 
effective temperature, 
is determined to be 0.428$\pm$0.003(stat.)$\pm$0.059(syst.).
An alternative description is provided by the weighted effective 
temperature, introduced to account for the differences in the 
temperature profile and muon flux as a function of zenith angle. 
Using the latter estimation of temperature, the coefficient  
is determined to be 
0.352$\pm$0.003(stat.)$\pm$0.046(syst.). \\
\end{abstract}

% muon and other elementary Particle Detectorsw
%Cosmic rays-astronominal observation,
%Scintillation Detectors,
% Dark Matter maybe not 95.55.Vj maybe not 95.85.Ry

\maketitle

\section{Introduction}
\label{Introduction}

It is well-known that the fluxes of cosmic ray muons observed in underground
detectors exhibit a seasonal variation.  The flux variations are
attributed to density variations in the atmosphere,
where
mesons from
primary cosmic ray interactions are themselves strongly
interacting 
%via the strong interaction 
or 
decaying via the weak interaction.  During
the summer, increases in the temperature cause the atmosphere to
expand, reducing the probability that a secondary meson will
interact. Consequently, the muon flux from weak meson decays increases.  
This variation in the muon flux has been observed, and correlated with 
temperature changes, by a number of experiments 
\cite{Barret:1952,Sherman:1954,Torino:1967,Hobart:1961,Baksan:1987,
Ambrosio:1997tc,Bouchta:1999,Bellini:2012te,Solvi:2009,Desiati:2011,
Osprey:2009ni},
including the MINOS measurement at the Far Detector (FD)
\cite{Adamson:2009zf}. 
These experiments measure $\alpha_{T}$,
the correlation coefficient between the muon flux and the 
atmospheric temperature. This coefficient varies as a function 
of overburden. The much shallower MINOS Near Detector, 
with 225 mwe overburden, is at a depth where $\alpha_{T}$ 
is expected to be rapidly changing as a function of overburden 
and has never before been accurately measured. 
In this paper, we make the first measurement of 
$\alpha_{T}$ in this important region. We also develop a novel 
formalism that takes into account the variation in atmospheric 
overburden, and hence the effective temperature, as a function 
of zenith angle. 

%There are three differences between our Near Detector (ND)
%analysis presented here and our FD analysis in
%Reference~\cite{Adamson:2009zf}
%to note:  1) The ND
%is located at the Fermi National Accelerator Laboratory (Fermilab),
%734 km from the MINOS FD.
%2) The ND is 
%situated under a flat overburden of approximately 225 meters 
%of water equivalent~(mwe), which is shallower than the Far
%Detector (2080 mwe).  It is at a depth where the correlation coefficient 
%between the muon flux and the atmospheric temperature %
%(defined in Eq.~(\ref{eq:alpha}) below)
%is expected to be rapidly 
%changing as a function of overburden and has not yet been measured. 
%3) An alternative approach to the standard analysis is presented,
%using formalism which takes into account the variation in atmospheric
%overburden and hence the effective temperature with zenith angle.  Results from the two analyses are compared.

Meson decays take place over a range of altitudes where the 
temperature is changing.  
It is customary to define an effective atmospheric temperature $T_{eff}$,
described in Section~\ref{sec:Theory}, where the muons originate.
The variation in the observed muon rate $R_{\mu}$ 
compared to the mean rate $\<R_\mu\>$
can be 
expressed in terms of a similar change in $T_{eff}$:
\begin{equation}
\frac{\Delta R_{\mu}}
{\<R_{\mu}\>}=\alpha_{T}\frac{\Delta T_{eff}}{\<T_{eff}\>}.
\label{eq:alpha}
\end{equation}
where $\<R_{\mu}\>$ 
%is the mean muon rate, and 
is equivalent to the 
rate for an effective atmospheric temperature $\<T_{eff}\>$. The 
magnitude of the temperature coefficient $\alpha_{T}$ depends upon 
the muon energy and hence
upon the depth of the detector. The parameter $\alpha_{T}$ is larger 
for detectors 
situated deeper underground because the muons originate from higher 
energy mesons which have increased lifetimes due to time dilation. 
This parameter is reduced 
for shallow detectors because as the atmospheric temperature increases 
the primary interaction occurs at a higher altitude increasing the 
probability that the muon will itself decay  prior 
to reaching the detector.  A measurement of the temperature coefficient
may be used to measure the K/$\pi$ ratio at energies beyond the
reach of current fixed target experiments \cite{Grashorn:2009ey}.
%The MINOS ND is situated at a depth where the correlation 
%coefficient $\alpha_{T}$ between the muon flux and the atmospheric 
%temperature is rapidly changing and has not yet been measured. 
Moreover 
the size of the detector, combined with its angular resolution, has 
allowed the first measurements of $\alpha_{T}$ as a function of the 
muon zenith angle. The data analyzed in this work consist
of single muon events recorded by the MINOS ND over the
six-year period between
June 1,~2006 and April 30, 2012.\\ 

The selection of the experimental data sets is presented in Sec.~II
below.  Section~\ref{sec:NearDetectorMuons} 
presents the measurement of the temperature 
coefficient $\alpha_{T}$ along with the theoretical prediction.  
In Sec.~\ref{sec:ZenithAngle} the dependence of 
$\alpha_{T}$ on the muon zenith angle is examined.  The
observations motivate the introduction of a new 
formula for the effective temperature which improves
upon the approximations
that are present in the standard analysis. Conclusions from the
analysis of this work are presented in Sec.~\ref{sec:Conclusion}.\\

\section{The Data Sets}
The measurement of the temperature coefficient $\alpha_{T}$ 
has been performed using muon data collected by the MINOS 
ND and temperature data provided by the European 
Center for Medium-Range Weather Forecasts (ECMWF) \cite{ECMWF}.

\subsection{MINOS Near Detector Muon Data}
\label{sec:MinosData}
%, with its long axis pointing \unit[26.5548]{$^\circ$} west of north
The \unit[0.98]{kton} MINOS ND \cite{Michael:2008bc} is 
a magnetized steel and scintillator sampling calorimeter designed to
measure neutrino interactions in the Fermilab NuMI beam \cite{bib:numi}.  It is
located at Fermilab. 
The detector, whose dimensions are 
\unit[3.8]{m}~height$\times$\unit[4.8]{m}~width$\times$\unit[16.6]{m}~length,
contains 282 vertical planes.  Each of the first 120 planes  
consists of a \unit[2.54]{cm} thick steel 
plane, a \unit[1]{cm} thick scintillator layer and a small air gap.  
The scintillator layers are composed of either 64 or 96 
scintillating strips, each 4.1~cm wide.
  In the latter 162 planes only one in five steel planes
have an attached scintillating layer.  The strips in neighboring planes 
are orthogonal to allow for three-dimensional track reconstruction. The 
scintillating strips are read out by 64-pixel multi-anode photo-multiplier 
tubes (PMT) \cite{Tagg:2004bu}. \\

Each PMT pixel is digitized continuously at 
\unit[53.1]{MHz}~(\unit[18.83]{ns}).  
For this analysis, a cosmic trigger was used ~\cite{Michael:2008bc};
the trigger was produced
when either four strips in five sequential planes, or when 
strips from any 20 planes, register a signal above the 1/3 photo-electron 
dynode threshold within \unit[151]{ns}. This trigger rate at the 
MINOS ND is approximately \unit[27]{Hz}. \\

\begin{figure}[thb]
  \begin{center}
    \includegraphics[width=0.48\textwidth]{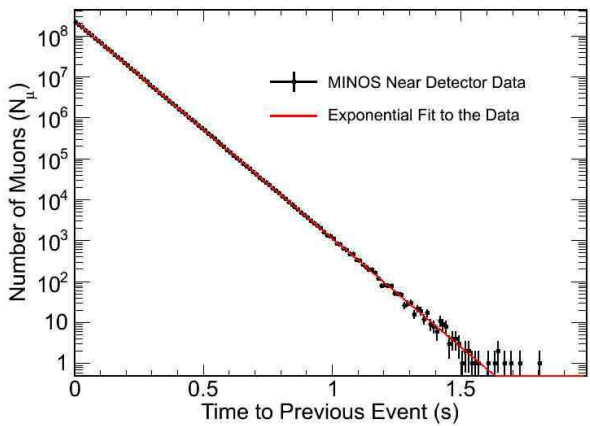}
  \end{center}
  \caption{Time between neighboring single atmospheric muon 
events in the MINOS ND. The data is well fitted to an exponential
distribution with a mean rate of \unit[12.2374$\pm$0.0003]{Hz}. }
  \label{fig:DeltaT}
\end{figure}

The atmospheric muon selection applied to the cosmic trigger 
data requires that the event contains one well 
reconstructed downward-going track that was collected during 
a period of good 
detector running conditions.  The requirements are the same as those used for
the MINOS analysis of the ND charge ratio \cite{Adamson:2010xf} up to the
charge sign quality selection in that analysis.  Comparison with Monte
Carlo shows that backgrounds and misreconstruction errors are negligible.
%That the event time distribution is well-behaved can be seen
%in Fig.~\ref{fig:DeltaT}, which shows the time 
%between neighboring  muon events.
Figure~\ref{fig:DeltaT} demonstrates the distribution of the time
between consecutive muon events is exponential, as expected.
In total over 1.53x10$^{9}$  single 
muon events have been selected with a mean 
rate of \unit[12.2374$\pm$0.0003]{Hz}.  The trigger rate above reflects
real muons, and the reduction is mostly due to the fact that the
scintillator coverage in the ND is smaller than the steel.
This geometry effect has no impact on the seasonal variation.

\subsection{Effective Temperature}
\label{sec:Theory}

The temperature as a function of atmospheric depth has been 
determined using the European Center for Medium-Range Weather 
Forecasts (ECMWF) atmospheric model \cite{ECMWF}.  The ECMWF procedure 
collates a number of different types of observations (e.g., 
surface, satellite, upper air sounding) at approximately 640 
locations around the globe; the data are contiguous both 
spatially and in time. The ECMWF global atmospheric model 
interpolates to a particular location assuming a smooth 
function of 1 degree latitude and longitude bins, and in 
varying elevation bins. For the MINOS ND  
at Fermilab, the model calculates atmospheric temperatures 
at 37 different, unevenly spaced pressure levels between 
\unit[1]{hPa} and \unit[1000]{hPa} at four times throughout the 24 hour day 
(0000h, 0600h, 1200h, 1800h.). An earlier version of the ECMWF 
model calculates temperatures at 21 pressure levels, and was 
used to help determine the sensitivity of the $\alpha_T$ fits.  By 
comparing the ECMWF temperature data with that of the 
Integrated Global Radiosonde Archive (IGRA) \cite{Durre:2006}, 
%(reference: I. Durre, R.S. Vose, and D. Wuertz, J. Clim. 19, 53 (2006)) 
it was determined that the uncertainties are 0.31 K.
As is reported in Ref.~\cite{Adamson:2009zf}, the systematic
uncertainty for this temperature model is estimated 
to be 0.2\%.

\begin{figure}[thb]
  \begin{center}
    \includegraphics[width=0.48\textwidth]{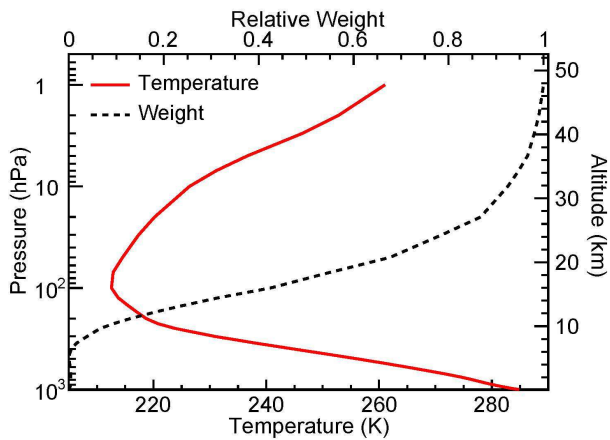}
  \end{center}
  \caption{The average temperature~(solid red line) \cite{ECMWF} 
and normalized weights $W(X)$~(blacked dashed line) as a function 
of pressure level at the MINOS ND site. The right 
vertical axis shows the altitude corresponding to a particular pressure. }
  \label{fig:Weight}
\end{figure}

The lack of data above 
a height corresponding to 
\unit[1]{hPa} does not affect the results of this 
analysis 
as the depth $X$ of the atmosphere above 
\unit[1]{hPa}~(\unit[1]{hPa} = \unit[1.019]{g/cm$^{2}$}) is 
insufficient to produce a statistically significant number 
of muons.  Since it is not possible to determine where in 
the atmosphere a particular muon originated, a 
single effective temperature is defined ~\cite{Grashorn:2008dis,
Grashorn:2009ey}, $T_{eff}$, which is the weighted $W(X)$ average based on 
the expected muon production spectrum 
\begin{equation}
\label{eq:teff}
T_{eff}=\frac{\int_{0}^{\infty}dX~ T(X)W(X)}{\int_{0}^{\infty}dX ~W(X)}.
\end{equation}
Since the temperature $T(X)$ is measured at 37 discrete depths,
a numerical integration is performed based on a quadratic interpolation
between temperature measurements.  
The atmospheric depth depends
on both $\pi$ and K decay, 
so 
$W(X)=W^{\pi}(X)+W^{K}(X)$ 
Figure~\ref{fig:Weight} shows the mean temperatures, averaged over 
the analysis period, and the normalized weight $W(X)$ 
as a function of atmospheric depth.

These weights are
\begin{equation}
W^{\pi~(K)}(X)\approx\frac{(1-X/\Lambda_{\pi~(K)}^{\prime})^{2}
e^{-X/\Lambda_{\pi~(K)}}A_{\pi~(K)}^{1}}{\gamma+(\gamma+1)B^{1}_{\pi~(K)}K(X)
(\frac{<E_{th} \cos \theta>}{\epsilon_{\pi~(K)}})^{2}}
\label{eq:teff3}
\end{equation}
where
\begin{equation}
K(X)=\frac{(1-X/\Lambda^{\prime}_{\pi~(K)})^{2}}
{(1-e^{-X/\Lambda^{\prime}_{\pi~(K)}})\Lambda_{\pi~(K)}^{\prime}/X}.
\label{eq:teff4}
\end{equation}
The attenuation lengths of the cosmic ray primary, pion and kaon are 
$\Lambda_{N}$, $\Lambda_{\pi}$ and $\Lambda_{K}$ respectively. 
$\Lambda_{\pi~(K)}^{\prime}$ is defined as 
$1/\Lambda_{\pi~(K)}^{\prime}=1/\Lambda_{N}-1/\Lambda_{\pi~(K)}$.  
The parameters $A_{\pi~(K)}^{1}$ account for inclusive 
meson production in the forward fragmentation region, the masses 
of mesons and muons and the muon spectral index 
$\gamma$~\cite{Grashorn:2008dis,
Grashorn:2009ey}.  The parameters $B_{\pi~(K)}^{1}$ reflect the relative 
attenuation of mesons in the atmosphere.  The critical energy of the 
mesons $\epsilon_{\pi~(K)}$ are the energies at which the probability of 
meson decay or interaction are equal.  $E_{th}$ is the minimum energy 
required for a muon to survive to a particular depth and $\theta$ is 
the zenith angle of the muon.  Apart from the value of 
$\<E_{th} \cos \theta\>$, 
which has a mean value of \unit[54]{GeV} at the MINOS ND, the 
values used for the parameters in Eq.~(\ref{eq:teff3}) and 
Eq.~(\ref{eq:teff4}) are the same 
as in Refs.~\cite{Adamson:2009zf,Bellini:2012te}.\\

\section{Data Analysis}
\label{sec:NearDetectorMuons}

Equation (\ref{eq:alpha}) states that the change in the observed muon 
rate is related to the change in the effective atmospheric temperature. 
In this section we will present the MINOS ND muon and ECMWF 
temperature data as a function of time.  The value of $\alpha_{T}$ is 
then determined by comparing the effective temperature determined from 
a single ECMWF temperature measurement to the corresponding six hours 
of MINOS muon data~(\unit[$\pm$3]{hours} on either side). The effect of 
surface pressure on the muon rate
was investigated and found to be 
small \cite{Sagisaka:1986bq,Motoki:2002nr}. It had no impact on the 
measurement of $\alpha_{T}$ and is therefore not considered further.

\subsection{Seasonal Variations}
\label{sec:SeasonalVariations}
\begin{figure}[thb]
%RawDataPlots/FitRatePlots.C
%An additional error of \unit[0.44]{K} has been added in quadrature as 
%the statistical flight-to-flight variation in the ECMWF data set. 
  \begin{center}
    \includegraphics[width=0.48\textwidth]
{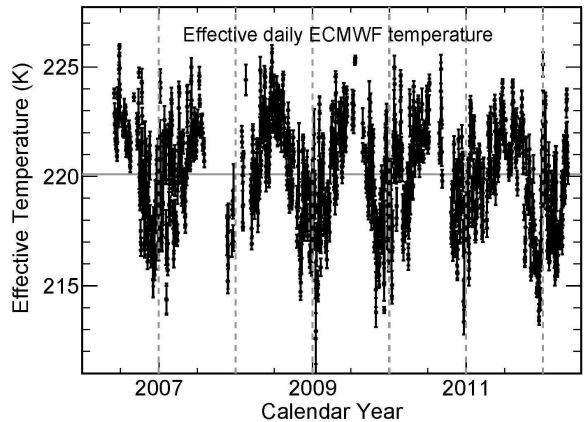}
  \end{center}
  \caption{Effective temperature as a function of time for the 
atmosphere directly above the MINOS ND. Each data point 
corresponds to one day of ECMWF data. The mean value is the average 
of the four ECMWF data points for that day. The $y$-axis errors are the 
standard deviation of those points.  The solid horizontal line is the 
mean effective temperature $\<T_{eff}\>$=\unit[220.1]{K}. The dashed 
vertical lines denote the start of new calendar years.}
  \label{fig:DailyTeff}
\end{figure}

\begin{figure}[thb]
  \begin{center}
    \includegraphics[width=0.48\textwidth]{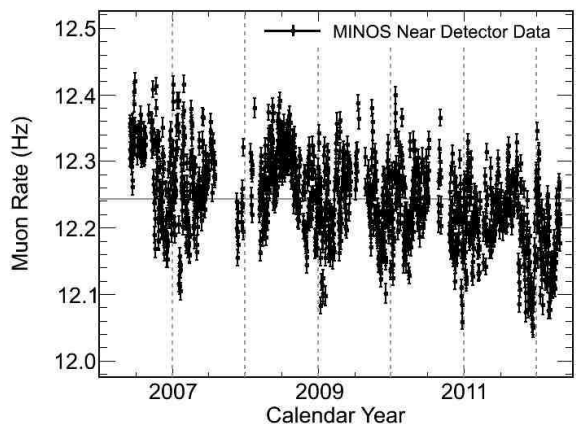}
  \end{center}
  \caption{The observed muon rate at the MINOS ND as a 
function of time. Each data point corresponds to one day of data. 
The horizontal line is the detector average of \unit[12.2458]{Hz}. 
The dashed vertical lines mark the start of new calendar years.}
  \label{fig:MuonRatePlot}
\end{figure}

Figure~\ref{fig:DailyTeff} displays the effective atmospheric temperature, 
as defined by Eq.~(\ref{eq:teff}),  directly above the MINOS ND
as a function of time. Figure~\ref{fig:MuonRatePlot} shows 
the observed muon rate at the MINOS ND as a function of 
time. The gaps in the data correspond to periods when the ND 
was not running or when the detector failed the data quality criteria. \\


Both the MINOS ND muon and effective temperature data 
have clear modulation signatures. The nominal modulation parameters 
were determined by fitting the data to an equation of the form
\begin{equation}
R(t)=R_{0}\left(1+A\cdot \textrm{cos}
\left[\frac{2\pi}{P}(t-t_{0})\right]\right),
\label{eq:Cosine}
\end{equation}
where $R_{0}$ is mean value, $A$ is the fractional modulation amplitude 
and $P$ is the period.  The time $t$ is the number of days elapsed 
since Jan.~1, 2010.  The phase $t_{0}$ is the first day at which the signal 
is at a maximum.  Fitting the MINOS ND muon data in 
Fig.~\ref{fig:MuonRatePlot} to Eq.~(\ref{eq:Cosine}) yields a mean 
rate of \unit[12.2458$\pm$0.0003]{Hz}, a period of 
\unit[367.8$\pm$0.4]{days} and a phase of \unit[200.9$\pm$0.8]{days}.  
Fitting the effective temperature data in Fig.~\ref{fig:DailyTeff} 
to Eq.~(\ref{eq:Cosine}) yields a mean value of \unit[220.1$\pm$0.2]{K}, 
a period of \unit[365.0$\pm$0.1]{days} and a phase of 
\unit[183.4$\pm$0.3]{days}.  As expected the modulation periods for 
both data sets are close to one year with the maxima occurring in 
the summer months. \\

\begin{figure}[thb]
  \begin{center}
    \includegraphics[width=0.48\textwidth]{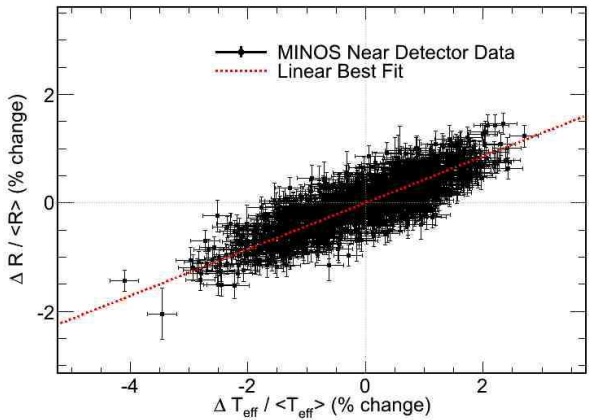}
  \end{center}
  \caption{Distribution of ${\Delta R}/{\<{R}\>}$ versus 
${\Delta T_{eff}}/{\<{T_{eff}}\>}$. Each data point corresponds to 
approximately 6 hours of MINOS ND data. The $y$-axis 
uncertainty is purely statistical. The $x$-axis uncertainty is 0.2\% and is 
the point-to-point variation in the ECMWF data. The best fit slope, 
equivalent to $\alpha_{T}$, is 0.465$\pm$0.003(stat.).  To reduce 
clutter, only every fifth data point is shown.}
  \label{fig:RateVsTime}
\end{figure}

Figure~\ref{fig:RateVsTime} shows the percentage change in the observed 
muon rate ${\Delta R}/{\<{R}\>}$ versus the per cent change in 
effective temperature ${\Delta T_{eff}}/{\<{T_{eff}}\>}$. The two 
data sets are strongly correlated with a correlation coefficient 
$\rho$=0.81. The best fit value for $\alpha_{T}$ is 0.465$\pm$0.003(stat.). \\
%The $\chi^{2}$ per degree of freedom is 12708/6367. 

The data in Fig.~\ref{fig:MuonRatePlot} indicate 
that the mean muon rate has decreased over the lifetime of the 
experiment.  The source of this small but apparently steady decrease
has not been conclusively identified.  
Three possible sources of this rate loss have been identified:
(i) solar cycle 
effects on the primary cosmic-ray rate, 
(ii) secular variations in the local magnetic field, and 
(iii) detector degradation effects. Since the effect seems 
larger for longer tracks than for shorter 
tracks, a detector degradation explanation is disfavored. The 
rate loss could possibly be reflected in the temperature
and represent a shortcoming of 
the temperature data. Biases and trends have been 
reported with ERA-Interim temperature data, 
most notably around 200-100 hPa  \cite{Simmons:2014}.
These have been attributed to warm biases in 
aircraft observations entering the data assimilation. 
However, these can only explain 10\% of the 
observed rate loss, as comparative temperature 
biases with Radiosonde data are \unit[0.1]{K} \cite{Dee:2009}.
Regardless of its causes, the effect
can be almost entirely removed 
by assuming a linear decline and refitting the data to obtain $\alpha_{T}$.
To do this, Equation~(\ref{eq:alpha}) can be modified to account for a 
rate loss by redefining $\<R_{\mu}\>$ as 
\begin{equation}
\<R_{\mu}^{t}\>~=~\<R_{\mu}^{0}\>\cdot\left(1-f\cdot\frac{t}{365.25}\right),
\label{eq:LossRate}
\end{equation}
where $f$ is the fractional loss rate, $t$ is the number of days 
since Jan.~1, 2010 and $\<R_{\mu}^{0}\>$ is the mean muon rate on that
date. 
The data were again fit, this time allowing for the mean muon rate to 
change as a function of time according to Eq.~(\ref{eq:LossRate}), 
and the best fit value of $\alpha_{T}$ was calculated to 
be 0.428$\pm$0.003(stat.).  This value comprises our result
using the standard definition of effective temperature.

\subsection{Systematic Uncertainties}
\label{sec:Systematics}
The systematic uncertainties on $\alpha_{T}$ can be loosely grouped 
into two sources, those derived from the analysis of the muon data, 
and those relating to the calculation of the effective temperature. 
This Section elaborates on the determination of these uncertainties 
whose magnitudes are given in Table \ref{tab:VerticalSystematics}. \\

\begin{table}[htb]
\begin{tabular}{lc}\hline
Systematic & $\alpha_{T}$ Uncertainty in  $\alpha_{T}$\\ \hline
Muon Direction & 0.017 \\
Rate loss Fit & 0.018 \\ 
Integration    & 0.023 \\
ECMWF Model    & 0.018 \\
Temp. Series   & 0.045 \\\hline
\multicolumn{2}{c}{T$_{eff}$ Calculation}\\ \hline
%R$_{K/\pi}$  & 0.0007 \\
%$\epsilon_{K}$& 0.000002\\
%$\epsilon_{\pi}$& 0.0006\\
%$\gamma$  &0.00025\\
\<E$_{th} \cos \theta$\> & 0.0023\\ \hline
Net Systematic & $\pm$0.059 \\ \hline
\end{tabular}
\caption{The systematic uncertainties associated with the nominal 
measurement of $\alpha_{T}$.}
\label{tab:VerticalSystematics}
\end{table}

The nominal effective temperature has been determined using 
the atmospheric temperature profile directly above the detector.  
However, the temperature profile will change as a function of 
latitude and longitude. This implies that the effective temperature, 
and therefore $\alpha_{T}$, is a function of the arrival direction of 
the muon.   The muon data were separated into northerly and 
southerly-going components, in order to maximize exposure 
to differences in the 
atmospheric temperature profiles.  A value of $\alpha_{T}$ (using 
the nominal $T_{eff}$) was calculated for each data set. The maximum 
difference from the nominal value, $\pm$0.017, is the systematic 
uncertainty due to the variability in the temperature profile.\\

The muon rate is clearly decreasing a small amount since
the beginning of the experiment, but the decrease
need not be linear as our fit assumes.  
The systematic uncertainty associated with decreasing event rate,
based upon the change implied by allowance for the fitted rate loss,
is estimated to be 
%50\% of the difference 
%between the value with and without the rate loss fit, 
$\pm$0.018.\\

For this analysis the two integrals in the definition of
T$_{eff}$ in Eq.~(\ref{eq:teff})
were evaluated using a quadratic interpolation technique. Multiple integration 
techniques were tested and the maximum difference from the employed method, 
$\pm$0.023, is the systematic uncertainty associated with the integration 
technique. To evaluate the uncertainty associated with the ECMWF 
temperature data itself, the  $\alpha_{T}$ parameter was re-evaluated
using an older 21 
pressure-level ECMWF model. Fitting only the data from the periods 
where the two models overlap, the best-fit values are found to differ
by $\pm$0.018.  This difference is taken to be
the uncertainty due to the ECMWF model.\\

\begin{figure}[thb]
  \begin{center}
    \includegraphics[width=0.48\textwidth]{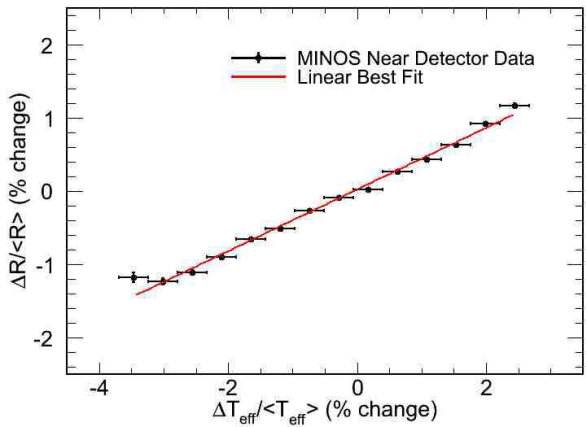}
  \end{center}
  \caption{The change in the observed muon rate versus the change 
in the effective temperature. In this Figure the muon rate has
been calculated as a function of effective temperature rather 
than on a point-to-point basis as in Fig.~\ref{fig:RateVsTime}.}
  \label{fig:RateVsTempSeries}
\end{figure}

The nominal value of $\alpha_{T}$ given in 
Section~\ref{sec:SeasonalVariations} was determined by comparing the 
muon rate measured over a six hour interval to the average effective 
temperature for that period. An alternative approach would be to 
calculate the mean muon rate for a given effective atmospheric 
temperature. The data have been grouped into 1~K bins in 
temperature (roughly twice the statistical uncertainty) and the 
muon rate determined as the total number of events divided by the 
total time for the data points that occur in that bin. 
Figure~\ref{fig:RateVsTempSeries} shows the per cent change in mean 
muon rate versus the per cent change in effective atmospheric 
temperature. The best fit value for $\alpha_{T}$ is 0.420$\pm$0.015(stat.). 
The deviation of this value  from the nominal value, $\pm$0.045, is the 
systematic uncertainty associated with the analysis technique.\\

Lastly, there is uncertainty in the parameters used to calculate 
the effective temperature. Of the parameters studied 
in \cite{Adamson:2009zf},  namely R$_{K/\pi}$, $\epsilon_{K}$, 
$\epsilon_{\pi}$, $\gamma$ and $\<{\rm E}_{th}$cos$\theta\>$, it was 
found that only $\<{\rm E}_{th}$cos$\theta\>$ =~(\unit[54]{GeV}$\pm$10\%) 
had a non-negligible impact, $\pm$0.0024, on the calculated 
value of $\alpha_{T}$.\\

In summary, the effective temperature 
coefficient $\alpha_{T}$ at the MINOS ND is determined to be 
0.428$\pm$0.003(stat.)$\pm$0.059(syst.).

\subsection{Theoretical Prediction}
\label{sec:Expectation}
The theoretical value of $\alpha_{T}$ can be expressed in terms 
of the differential muon intensity I$_{\mu}$ as \cite{Barret:1952}:
\begin{equation}
\alpha_{T}=-\frac{E_{th}}{I_{\mu}^{0}}
\frac{\partial I_{\mu}}{\partial E_{th}}-\gamma.
\end{equation}
Performing the differentiation yields~\cite{Barret:1952,Grashorn:2009ey}
\begin{equation}
\alpha_{T}=\frac{1}{D_{\pi}}
\frac{1/\epsilon_{K}+A_{K}^{1}(D_{\pi}/D_{K})^{2}/\epsilon_{\pi}}
{1/\epsilon_{K}+A_{K}^{1}(D_{\pi}/D_{K})/\epsilon_{\pi}}-\delta
\label{eq:TheoryAlpha}
\end{equation}
where
\begin{equation}
D_{\pi/K}=\frac{\gamma}{\gamma+1}
\frac{\epsilon_{\pi/K}}{1.1 \<E_{th}\textrm{cos}\theta\>}+1
\end{equation}
and the correction for muon decay $\delta$ is
\begin{equation}  
\delta=1.0336\frac{\gamma}{\gamma+1}\textrm{ln}
\left(\frac{8.5833}{\textrm{cos}\theta}\right)
\frac{1}{\<E_{th}\textrm{cos}\theta\>}.
\end{equation}
A Monte Carlo simulation was used to determine the theoretically 
expected value of $\alpha_{T}$.  A muon energy and cos$\theta$ 
were chosen randomly from the differential muon intensity 
spectrum~\cite{Gaisser:1990vg}. The muon was then randomly 
assigned an azimuthal angle $\phi$.  
%If the muon energy was 
%greater than the threshold energy required to reach the 
%MINOS ND. 
The threshold energy for a particular 
direction in $\theta$ and $\phi$ was determined using the MINOS 
overburden; details are given in Ref.~\cite{Adamson:2010xf}. 
The $\alpha_{T}$ parameter
was calculated using Eq.~(\ref{eq:TheoryAlpha}).  This process 
was repeated to obtain an $\alpha_{T}$ distribution generated 
from 10,000 successful muon events.  The theoretical value of 
$\alpha_{T}$ is the mean of this distribution and is equal to 
0.390$\pm$0.004(stat.).  The theoretical value of $\alpha_{T}$ 
has a systematic uncertainty due to the uncertainties in the 
parameters used to evaluate Eq.~(\ref{eq:TheoryAlpha}).  
Table \ref{tab:TheoryError} gives the $\pm$1$\sigma$ 
uncertainties with the respective parameters.\\

\begin{table}[htb]
\begin{tabular}{lc}\hline
Systematic & Uncertainty on $\alpha_{T}$ \\ \hline
R$_{K/\pi}$~=~0.149$\pm$0.06                        & 0.011 \\
$\epsilon_{K}$~=~\unit[0.850$\pm$0.014]{TeV}        & 0.00016\\
$\epsilon_{\pi}$~=~\unit[0.115$\pm$0.003]{TeV}      & 0.00567\\
$\gamma$~=~\unit[1.7$\pm$0.1]{}                     & 0.00556\\
$\<E_{th}\textrm{cos}\theta\>$~=~\unit[54]{GeV}$\pm$10\% & 0.0243\\ \hline
Net Systematic                                    & $\pm$0.028 \\ \hline
\end{tabular}
\caption{The $\pm$1$\sigma$ systematic uncertainties on the 
theoretical value of $\alpha_{T}$ at the MINOS ND.}
\label{tab:TheoryError}
\end{table}

The measured value of $\alpha_{T}^{exp}$=0.428$\pm$0.003(stat.)
$\pm$0.059(syst.) is larger than, but consistent, with the theoretical 
prediction of $\alpha_{T}^{theory}$=0.390$\pm$0.004(stat.)
$\pm$0.028(syst.).\\

\section{Zenith Angle Analysis}
\label{sec:ZenithAngle}
The measurement of $\alpha_{T}$ in the preceding Section assumes that 
the variation in the muon rates at all zenith angles only depends 
upon the vertical effective temperature~(Eq.~(\ref{eq:teff})). However, 
cosmic ray primaries with large zenith angles interact higher in the 
atmosphere where the temperature fluctuations are larger. Consequently, 
the variation in the muon rates should increase as a function of the 
zenith angle and, with no redefinition of the effective temperature, 
the measured values of $\alpha_{T}$ should increase as well.  In this 
section we will calculate $\alpha_{T}$ as a function of zenith angle 
using both the vertical and angular effective temperatures. \\

In addition to the selection criteria outlined in 
Sec.~\ref{sec:MinosData} the angular resolution of the 
muon tracks is required 
to be better than 5$^{\circ}$.  
So as to not change the underlying E$_{th}\textrm{cos}\theta$ 
distribution of the muons,
the changes in event selection were kept to a 
minimum.
The value of $\alpha_{T}$ was determined for this
resolution-enhanced data sample 
to be statistically consistent with the nominal value, 0.428.\\

\begin{figure}[thb]
  \begin{center}
    \includegraphics[width=0.48\textwidth]{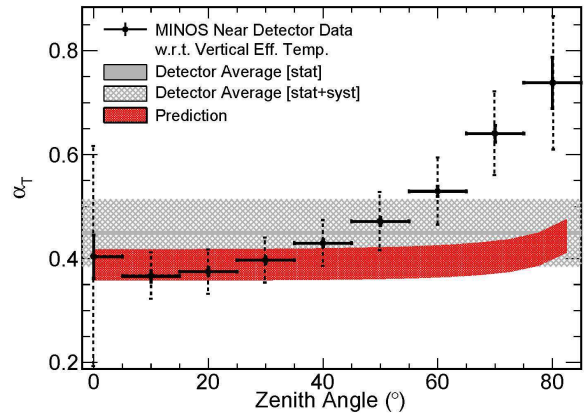}
  \end{center}
  \caption{The parameter $\alpha_{T}$ as a function of zenith angle 
when $T_{eff}$ is calculated using Eq.~(\ref{eq:teff}).  
The inner error bars on the data points are statistical, 
the outer error bars include both statistical and 
systematic uncertainties.  The gray band is the detector average, 
and the red-band is the theoretical prediction.  The systematic
difference of the data from the average suggests that the vertical 
effective temperature is inadequate for data over a range of zenith
angles.}
  \label{fig:AlphaTVertical}
\end{figure}

Figure~\ref{fig:AlphaTVertical} gives the measured $\alpha_{T}$ as 
a function of zenith angle when $T_{eff}$ is calculated using 
Eq.~(\ref{eq:teff}).  The data are grouped into, and the values 
of $\alpha_{T}$ calculated for, nine zenith angle bins. The first bin is 
from 0-5$^{\circ}$, and the remaining 8 bins each cover the 
next 10$^{\circ}$ increments.  The theoretical prediction as a function 
of zenith angle is calculated using the Monte Carlo method outlined in 
Sec.~\ref{sec:Expectation} but averaged instead over the zenith angle 
bins.  It should be noted that the theoretical value 
of $\alpha_{T}$ is independent of the atmospheric temperature and is 
therefore not affected by our zenith angle corrections.
Not surprisingly the measured value of $\alpha_{T}$ increases with zenith angle and 
does so more rapidly than the theoretical prediction. \\

\begin{figure}[thb]
  \begin{center}
    \includegraphics[width=0.48\textwidth]{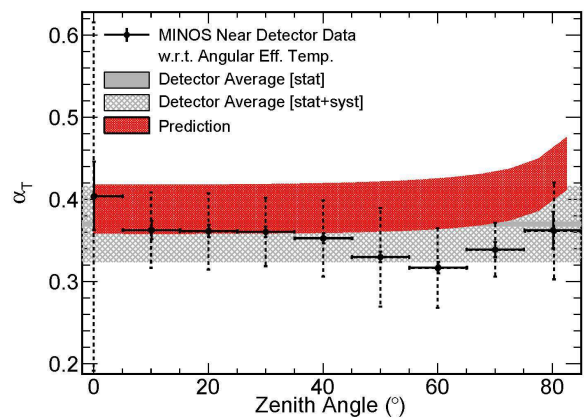}
  \end{center}
  \caption{The $\alpha_{T}$ parameter as a function of zenith angle 
when $T_{eff}(\theta)$ is calculated using Eq.~(\ref{eq:teff200}). The 
inner error bars on the data points are statistical, 
the outer error bars include both statistical and 
systematic uncertainties. The gray band is the detector average 
calculated using $T_{eff}^{\mathrm{weight}}$ as defined 
in Eq.~(\ref{eq:WeightedTemp}). The red band is the theoretical 
prediction and is the same as in Fig.~\ref{fig:AlphaTVertical}. }
  \label{fig:AlphaTeffective}
\end{figure}

Equation~(\ref{eq:teff}) was modified to account for the increased height of 
the primary cosmic ray interaction at larger zenith angles. The angular 
effective temperature for a particular zenith angle $\theta$ is simply
\begin{equation}
T_{eff}=\frac{\int_{0}^{\infty}d(X/\cos \theta)~T(X)~W(X)}
{\int_{0}^{\infty}d(X/\cos \theta)~W(X)}.
%\\
%T_{eff}(\theta)\approx\frac{\sum_{1}^{37}T_{n}\Delta(\frac{X_{n}}
%{\textrm{cos}\theta})(W_{n}^{\pi}+W_{n}^{K})}{\sum_{1}^{37}\Delta(\frac{X_{n}}{\textrm{cos}\theta})(W_{n}^{\pi}+W_{n}^{K})}.
\label{eq:teff200}
\end{equation}
The formulae for the weights $W(X)$ are 
unchanged from Eq.~(\ref{eq:teff3}), only the 
depth~($X\rightarrow X/\textrm{cos}\theta$) and  
$E_{th}\textrm{cos}\theta$ arguments change with zenith angle. 
The $1/\cos\theta$ terms in the denominator and numerator do cancel 
but have been left in for completeness. Figure~\ref{fig:AlphaTeffective} 
shows $\alpha_{T}$ as a function of zenith angle when the effective 
temperature $T_{eff}(\theta)$ has been calculated using 
Eq.~(\ref{eq:teff200}). The Figure shows that the values of 
$\alpha_{T}$ calculated in this manner are now consistent with 
the theoretical prediction. \\

\begin{figure}[thb]
  \begin{center}
    \includegraphics[width=0.48\textwidth]{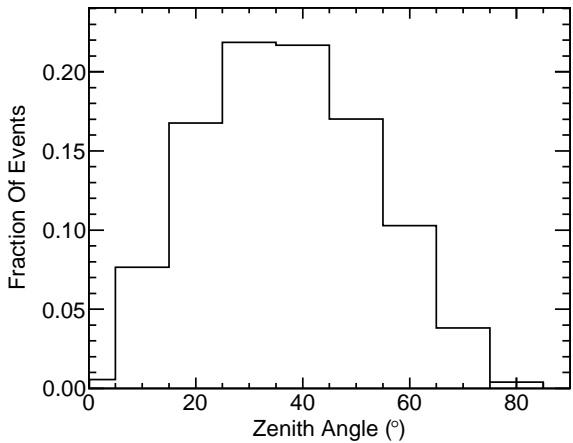}
  \end{center}
  \caption{Reconstructed zenith angle distribution for single 
muon events observed in the MINOS ND. 
}
  \label{fig:ZenithAngle}
\end{figure}

To determine a single value of $\alpha_{T}$ for the MINOS ND,
a single measure of temperature is initially defined
using 
Eq.~(\ref{eq:teff}), as a weighted average based upon the observed 
muon angular distribution. The weighted angular effective 
temperature is then defined as
\begin{equation}
T_{\mathrm{eff}}^{\mathrm{weight}}
=
\sum_{i=1}^{M}F_{i}\cdot T_{eff}(\theta_{i}),
\label{eq:WeightedTemp}
\end{equation}
where $M$ is the number of zenith angle bins. 
$T_{eff}$($\theta_{i}$) is the angular effective temperature 
in bin $i$. $F_{i}$ is the fraction of muons occurring in that bin, 
the distribution of which is shown in Fig.~\ref{fig:ZenithAngle}.\\

Using the weighted effective temperature defined 
in Eq.~(\ref{eq:WeightedTemp}), and 
repeating the data analysis and systematic calculations as described in 
Sec,~\ref{sec:NearDetectorMuons}, the weighted effective temperature 
coefficient $\alpha_{T}^{\mathrm{weight}}$ at the MINOS ND is found to be:
\begin{equation}
\alpha_{T}^{\mathrm{weight}}=0.352 \pm 0.003(\mathrm{stat.}) \pm 0.046(\mathrm{syst.}).
\end{equation}
The magnitudes of the individual systematic uncertainties are given in 
Table \ref{tab:WeightedSystematics}.  This result is consistent 
with the theoretical prediction of 
$\alpha_{T}^{theory}$ = 0.390$\pm$0.004(stat.)$\pm$0.028(syst.).   \\
\begin{table}[htb]
\begin{tabular}{lc}\hline
Systematic & Uncertainty on $\alpha_{T}^{\mathrm{weight}}$\\ \hline
Muon Direction & 0.020 \\
Rate loss Fit & 0.017 \\ 
Integration    & 0.033 \\
ECMWF Model    & 0.014\\
Temp. Series   & 0.011\\\hline
\multicolumn{2}{c}{T$_{eff}$ Calculation}\\ \hline
R$_{K/\pi}$  & 0.00186 \\
%$\epsilon_{K}$& 0.000006\\
%$\epsilon_{\pi}$& 0.0009\\
%$\gamma$  &0.00038\\
$\<E_{th} \cos \theta\>$ & 0.0036\\ \hline
Net Systematic & $\pm$0.046 \\ \hline
\end{tabular}
\caption{The systematic uncertainties associated with the 
measurement of $\alpha_{T}^{\mathrm{weight}}$. }
\label{tab:WeightedSystematics}
\end{table}

\begin{figure}[thb]
  \begin{center}
    \includegraphics[width=0.48\textwidth]{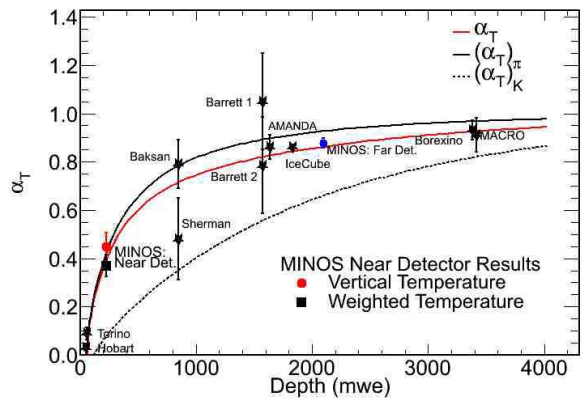}
  \end{center}
  \caption{The parameter $\alpha_{T}$ as a function of detector depth. The two 
measurements of this analysis are shown in black.  The middle red 
line is the theoretical prediction, Eq.~(\ref{eq:TheoryAlpha}), for a 
detector at depth in ($\rho$ = 2.65 gm/cm$^3$)
standard rock. The top~(bottom) line is the expected 
value of $\alpha_{T}$ assuming a pion-only (kaon-only) model determined by 
setting $A_{K}^{1}=0$~($A_{K}^{1}\rightarrow\infty$) \cite{Grashorn:2008dis}. }
  \label{fig:AlphaT}
\end{figure}

Figure~\ref{fig:AlphaT} shows the new MINOS ND results and 
all the known measured 
values of $\alpha_{T}$ as a function of detector 
depth. The Figure includes results from 
Barret 1,2~\cite{Barret:1952}, AMANDA~\cite{Bouchta:1999}, 
ICECUBE~\cite{Desiati:2011}, 
MACRO~\cite{Ambrosio:1997tc}, Torino~\cite{Torino:1967}, 
Hobart~\cite{Hobart:1961}, Sherman~\cite{Sherman:1954}, 
Baksan~\cite{Baksan:1987}, Borexino~\cite{Bellini:2012te} 
and the MINOS FD~\cite{Adamson:2009zf}. The 
data are fully consistent with the prediction that 
$\alpha_{T}$ increases with 
detector depth~(equivalent to increasing values of 
$E_{th}\textrm{cos}\theta$) and asymptotically approaches 
unity for very large detector depths.\\

\section{Conclusion}
\label{sec:Conclusion}
A measurement of the effective temperature coefficient 
$\alpha_{T}$ has been performed using nearly six years of MINOS ND data. 
The value of this coefficient is determined to be
\begin{equation}
\alpha_{T}=0.428\pm0.003(\mathrm{stat.})\pm0.059(\mathrm{syst.}).
\end{equation}
Additionally, a method that improves upon the conventional
approach to determination of 
$\alpha_{T}$ using an underground detector of
large angular acceptance has been demonstrated in this work.
The improvement is achieved by 
accounting for the variance in the modulation 
of muon rate as a function of zenith angle. 
A weighting of the effective temperature as a function 
of zenith angle based on the relative flux of muons 
improves consistency and gives: 
\begin{equation}
\alpha_{T}^{\mathrm{weight}}=0.352\pm
0.003(\mathrm{stat.})\pm0.046(\mathrm{syst.}).
\end{equation}
The zenith angle acceptance of an underground detector depends
on both the geometry of the detector and the geometry of the
overburden.   
The correction for zenith angle in
the determination of $\alpha_{T}$ is relatively more important
for detectors which have a vertically concave overburden, 
since these experience higher fluxes 
at large zenith angles.  However the zenith angle correction is
also important for detectors at
depths less than 1000 mwe where $\alpha_{T}$ is rapidly changing, as
shown in Fig.~\ref{fig:AlphaT}.
\section{Acknowledgments}
\label{sec:Acknowledgements} 
This work was supported by the US DOE, the United Kingdom STFC, 
the US NSF, the state and University of Minnesota, and Brazil's 
FAPESP, CNPq and CAPES. We are grateful to 
the personnel of Fermilab for their contributions to the 
experiment. 

\bibliography{SeasonalPaperOne}

\end{document}